\documentclass[aps,prd,twocolumn,showpacs,groupedaddress,preprintnumbers]{revtex4-1}

\begin{document}

\preprint{BROWN-HET-1606, WITS-CTP-56}

\title{AdS$_4$/CFT$_3$ Construction from Collective Fields}


\author{Robert de Mello Koch$^1$}
\email[]{Email: robert@neo.phys.wits.ac.za}
\author{Antal Jevicki$^2$}
\email[]{Email: antal\_jevicki@brown.edu}
\author{Kewang Jin$^2$}
\email[]{Email: kewang\_jin@brown.edu}
\author{Jo\~ao P. Rodrigues$^1$}
\email[]{Email: joao.rodrigues@wits.ac.za}
\affiliation{$^1$ National Institute for Theoretical Physics, \\
School of Physics and Centre for Theoretical Physics, \\ 
University of the Witwatersrand, Wits, 2050, South Africa \\
$^2$ Department of Physics, Brown University, Providence, RI 02912, USA}



\begin{abstract}

We pursue the construction of higher-spin theory in AdS$_4$ from CFT$_3$ of the O(N) vector model in terms of canonical collective fields. In null-plane quantization an exact map is established between the two spaces. The coordinates of the AdS$_4$ space-time are generated from the collective coordinates of the bi-local field. This, in the light-cone gauge, provides an exact one-to-one reconstruction of bulk AdS$_4$ space-time and higher-spin fields.

\end{abstract}


\maketitle


\section{Introduction}

The AdS/CFT correspondence \cite{Maldacena:1997re,Gubser:1998bc,Witten:1998qj} represents a very important tool in gauge and string theories. It gives a concrete, analytical procedure for the more general Gauge/String(Gravity) duality. The correspondence is characterized by conjectured emerging dimensions of space-time (in ${\cal N}=4$ Super Yang-Mills theory the $D=10$ of the string in $AdS_5 \times S^5$ background emerges). While the main understanding of the duality itself is provided by 't Hooft's large $N$ expansion (which establishes $1/N$ as the string coupling constant) the origin of the extra spatial dimension is less clearly understood, one speaks of them as being holographic and have a relationship (in the case of radial AdS dimension) with renormalization group scaling parameters.

One framework for analytical understanding of the Large $N$ limit in general introduced several decades ago \cite{Jevicki:1979mb} is based on the notion of collective fields. They capture the relevant degrees of freedom and a general method for describing their effective dynamics both at the Hamiltonian and Lagrangian level was given. This approach has been successful in analytical treatment as well as in exhibiting the relevant physics in various model theories. In the $c=1$ matrix model collective dynamics naturally led to (one) extra dimension relevant in establishing the model as a 2D non-critical string theory \cite{Das:1990kaa}. It has re-emerged in the sub-dynamics of the ${\cal N}=4$ Yang-Mills problem in the 1/2 BPS sub-sector. Through certain matrix model truncations (of ${\cal N}=4$ Yang-Mills theory) the construction \cite{deMelloKoch:2002nq} of dual string theory Hamiltonian was attempted.

For further understanding of this mechanism it is useful to concentrate on exactly solvable theories. The simplest field theory model for which one can build the AdS/CFT correspondence is that of $N$-component vector theory. It was originally pointed out by Klebanov and Polyakov \cite{Klebanov:2002ja} that the conformal fixed points of the theory are naturally described in four dimensional AdS space-time. More specifically it was established that it is a particular higher-spin theory of Vasiliev \cite{Vasiliev:1995dn} that emerges in the large $N$ limit. (In a series of works dating back to the 80's Vasiliev and collaborators have succeeded in constructing a remarkable theory providing interactions of a sequence of higher-spins in AdS. This (gauge) theory successfully extends a free theory \cite{Mikhailov:2002bp} obtained by Fronsdal \cite{Fronsdal:1978rb}.) An impressive comparison of three-point boundary correlators was performed recently by Giombi and Yin \cite{Giombi:2009wh}. For other relevant work, see \cite{HaggiMani:2000ru,Witten,Sezgin:2002rt,Leonhardt:2002ta}.

The relevance of collective fields for higher-spin holography was discussed by Das and one of the present authors \cite{Das:2003vw}. The framework of covariant bi-local collective fields was employed and it was shown that they decompose into an infinite sequence of integer spin fields in one extra dimension. The present paper sharpens this picture concentrating on the canonical formulation with the goal of establishing the correspondence directly at the Hamiltonian level. It will be advantageous to work in null-plane quantization, since it gives a physical description of higher-spin gauge theory. In this framework, we will produce an exact one-to-one map between (collective) coordinates of the large $N$ field and the AdS$_4$ coordinates of the higher-spin theory. It is shown how collective fields provide a construction of bulk (rather than boundary) fields of the AdS theory. In particular it is demonstrated that all the bulk AdS space-time transformation symmetries are recovered from transformations of the bi-local collective field. Outline of an exact map of the full interacting theory is given. 

This paper is organized as follows. In section II, we discuss the difference between collective fields and conformal currents which have been the main tool of earlier AdS/CFT comparisons. In section III, we summarize the form of the exact collective field Hamiltonian, discuss its expansion in $1/N$ as a coupling constant. Realization of the conformal symmetries and the quadratic approximation is studied in section IV. We establish a one-to-one map with the transformations of higher-spin theory in AdS$_4$ background in section V. Discussion of results and of further topics is done in the Conclusion.

\section{Collective vs conformal fields}

The basis of the holographic map is in a (complete) set of primary operators of the SO(2,d) group. They are built as composite operators from the basic fields of the theory and obey current conservation once the field equations are used. They are used as sources at the boundary and their correlators are then shown to be in agreement with the AdS amplitudes projected to the boundary of AdS space. The $N$-component vector model field theory with the Lagrangian
\begin{equation}
L=\int d^d x {1 \over 2}(\partial_\mu \phi^a)(\partial^\mu \phi^a)+v(\phi \cdot \phi), \quad a=1,...,N
\end{equation}
possesses two critical points: the UV fixed point at zero value of the coupling and an IR fixed point at nonzero coupling. For the UV case corresponding to the free theory where the potential $v=0$, a full set of conformal currents is explicitly given by \cite{Giombi:2009wh}
\begin{eqnarray}
{\cal O}(\vec{x},\vec{\epsilon})&=&\phi^a(x-\epsilon)\sum_{n=0}^\infty {1 \over (2n)!}\bigl(2\epsilon^2 \overleftarrow{\partial_x} \cdot \overrightarrow{\partial_x} \cr
& &-4(\epsilon \cdot \overleftarrow{\partial_x})(\epsilon \cdot \overrightarrow{\partial_x})\bigr)^n \phi^a(x+\epsilon)
\end{eqnarray}
where $\vec{\epsilon}$ is a null polarization vector $\vec{\epsilon}^{~2}=0$. These currents are conserved and in the holographic scheme of GKP-W \cite{Gubser:1998bc,Witten:1998qj} their correlators are compared with the AdS boundary amplitudes.

Collective fields for large $N$ theories are introduced in a very different manner. They are to represent a (complete) set of invariants under the $O(N)$ or $U(N)$ (gauge) symmetry group. The meaning of completeness is established in two not unrelated ways. First one has completeness in group theoretic terms, namely that any other invariant can be expressed in terms of them. Second is the requirement of closure under (quantum) equations of motion. This leads to the most important fact, namely that they provide a complete dynamical description \cite{Jevicki:1979mb} of the large $N$ theory where $1/N$ is seen to emerge as the  natural expansion parameter. 

In the $O(N)$ vector model one simply has the bi-local collective field
\begin{equation}
\Psi(x^\mu,y^\mu)=\sum_{a=1}^N \phi^a (x) \cdot \phi^a (y)
\end{equation}
in the covariant formalism \cite{Das:2003vw}. It is the case for the $O(N)$ model, and also more generally that the set of collective fields is actually over-complete. This property has significant implications on the emerging space-time, when implemented it naturally leads to space-time cutoffs and ultimately non-commutativity.

As far as the relationship between the conformal and collective fields we have the following. Clearly any conformal field is contained in the collective (bi-local) field, one has a prescription with derivatives given above. But the converse is not true, collective fields represent a more general set. This property will have important implications on the bulk vs boundary description of the theory. It has already seen in approximate manner \cite{Das:2003vw} that the relative coordinate in the bi-local field into angles generating a sequence of spins and the radial part which plays the role of an extra dimension. What prevented a precise identification however was the fact that higher-spin is a gauge theory, whose dynamical form depends on the gauge chosen. Consequently for establishing a precise one-to-one map, one has to bring both theories to the same gauge. This will be accomplished in the present work in a canonical description.

The canonical formalism for collective fields is based (in equal-time quantization) on the observables
\begin{equation}
\Psi(t;\vec{x},\vec{y})=\sum_{a} \phi^a (t,\vec{x}) \cdot \phi^a (t,\vec{y}) \equiv \Psi_{xy}
\end{equation}
which are local in time but bi-local in $d-1$ dimensional space. These observables (collective fields) are characterized by the fact that they represent a complete set of $O(N)$ invariant canonical variables (obtained through scalar product). To deduce the dynamics obeyed by these fields, one performs an operator change of variables \cite{Jevicki:1979mb} from $\phi^a(t,\vec{x})$ to the bi-local field $\Psi(t;\vec{x},\vec{y})$ using the chain rule
\begin{equation}
{\delta \over \delta \phi(\vec{x})}={\delta \Psi(\vec{y},\vec{z}) \over \delta \phi(\vec{x})}{\delta \over \delta \Psi(\vec{y},\vec{z})}.
\end{equation}
Starting from the canonical Hamiltonian
$$
H=\int \Bigl( -{1 \over 2}{\delta \over \delta \phi^a(\vec{x})}{\delta \over \delta \phi^a(\vec{x})} + {1 \over 2}\bigtriangledown_x \phi^a \bigtriangledown_x \phi^a + v(\phi \cdot \phi) \Bigr) d \vec{x},
$$
one deduces an equivalent representation in terms collective variables 
\begin{eqnarray}
H&=&2{\rm Tr}(\Pi \Psi \Pi)+{N^2 \over 8} {\rm Tr} \Psi^{-1} + \int d\vec{x} v(\Psi(\tilde{x},\tilde{y}) \vert_{\tilde{x}=\tilde{y}}) \cr
& &+{1 \over 2}\int d\vec{x} [-\bigtriangledown_x^2 \Psi(\tilde{x},\tilde{y}) \vert_{\tilde{x}=\tilde{y}}]+\Delta V
\label{Hamcol}
\end{eqnarray}
where we have the conjugate momentum denoted by
\begin{equation}
\Pi(\vec{x},\vec{y})=-i {\delta \over \delta \Psi(\vec{x},\vec{y})}
\end{equation}
and $\Delta V$ summarizes ordering terms which are lower order in $1/N$ 
$$
\Delta V=-{N \over 2}\Bigl( \int dx \delta(0) \Bigr){\rm Tr}\Psi^{-1}+{1 \over 2}\Bigl( \int dx \delta(0) \Bigr)^2 {\rm Tr} \Psi^{-1}.
$$
The product of two bi-local fields is defined by
\begin{equation}
AB=\int d\vec{y} A(\vec{x},\vec{y})B(\vec{y},\vec{z})
\end{equation}
and the trace of a bi-local field means
\begin{equation}
{\rm Tr}(A)=\int d\vec{x} A(\vec{x},\vec{x}).
\end{equation}
For more details on this representation, including the fact that it generates correctly the large $N$ Schwinger-Dyson equations, the reader should consult Refs. \cite{Jevicki:1979mb,Jevicki:1983hb}.

\section{Expansion}

The main feature of the collective  representation in terms of the Hamiltonian (\ref{Hamcol}) is that it can be expanded in series of $1/N$ with an infinite number of polynomial vertices to generate systematically the $1/N$ expansion. This is seen by a simple rescaling of field variables: $\Psi \to N \Psi, \Pi \to \Pi/N$ whereby $N$ factorizes in front of the action. The terms in $\Delta V$ are seen to be of lower order, consequently they provide counter-terms in the systematic $1/N$ expansion.

To generate the expansion, one first evaluates the static large $N$ background $\psi_0(\vec{x},\vec{y})$ obtained from the time-independent equations of motion
\begin{equation}
{\partial V \over \partial \Psi(\vec{x},\vec{y})}=0,
\end{equation}
where we have set $v=0$ and the effective potential reads
\begin{equation}
V={1 \over 8} {\rm Tr} \Psi^{-1}+{1 \over 2}\int d\vec{x} [-\bigtriangledown_x^2 \Psi(\tilde{x},\tilde{y}) \vert_{\tilde{x}=\tilde{y}}].
\end{equation}
One performs a shift 
\begin{equation}
\Psi=\psi_0 + {1 \over \sqrt{N}} \eta, \quad \Pi = \sqrt{N} \pi
\label{shift}
\end{equation}
generating an infinite sequence of vertices
\begin{equation}
{\rm Tr} \Psi^{-1} = {\rm Tr} \psi_0^{-1} + \sum_{n=1}^{\infty} {(-1)^n \over N^ {n\over 2}} {\rm Tr}  (\psi_0 (\eta \psi_0)^n).
\label{expansion}
\end{equation}
The quadratic and cubic terms in the Hamiltonian are seen to be given by  
\begin{eqnarray}
H^{(2)}&=&2 {\rm Tr}(\pi \psi_0 \pi)+{1 \over 8} {\rm Tr} (\psi_0 \eta \psi_0 \eta \psi_0), \\
H^{(3)}&=&{2 \over \sqrt{N}} {\rm Tr}(\pi \eta \pi) - {1 \over 8\sqrt{N}} {\rm Tr} (\psi_0 \eta \psi_0 \eta \psi_0 \eta \psi_0).
\end{eqnarray}
The higher order vertices are obtained directly from the expansion (\ref{expansion}).

We now discuss the evaluation of the spectrum which follows from diagonalization of $H^{(2)}$. In doing this we follow closely \cite{Jevicki:1983hb}. Using a Fourier transform
\begin{equation}
\psi_{xy}^0=\int d\vec{k} e^{i\vec{k}\cdot(\vec{x}-\vec{y})} \psi_k^0,
\end{equation}
with
\begin{equation}
\psi_k^0= {1 \over 2 \sqrt{{\vec{k}}^2}},
\end{equation}
and for the fields
\begin{eqnarray}
\eta_{xy} &\equiv& \int d\vec{k}_1 d\vec{k}_2 e^{-i\vec{k}_1 \cdot \vec{x}} e^{+i\vec{k}_2 \cdot \vec{y}} \eta_{k_1 k_2}, \\
\pi_{xy} &\equiv& \int d\vec{k}_1 d\vec{k}_2 e^{+i\vec{k}_1 \cdot \vec{x}} e^{-i\vec{k}_2 \cdot \vec{y}} \pi_{k_1 k_2},
\end{eqnarray}
the quadratic Hamiltonian now becomes 
\begin{eqnarray}
&&H^{(2)}=2 \int d\vec{k}_1 d\vec{k}_2 \psi^0_{k_1} \pi_{k_1 k_2} \pi_{k_1 k_2} \cr
&&+{1 \over 16} \int d\vec{k}_1 d\vec{k}_2 \eta_{k_1 k_2} ({\psi^0}_{k_1}^{-2} {\psi^0}_{k_2}^{-1} + {\psi^0}_{k_2}^{-2} {\psi^0}_{k_1}^{-1}) \eta_{k_1 k_2}. \nonumber
\end{eqnarray}
Redefining 
\begin{equation}
\pi_{k_1 k_2} \to {1 \over 2} {\psi^0}_{k_1}^{-1/2} \pi_{k_1 k_2} \quad \eta_{k_1 k_2} \to 2 {\psi^0}_{k_1}^{+1/2} \eta_{k_1 k_2}
\end{equation}
one has the quadratic Hamiltonian
\begin{eqnarray}
&&H^{(2)}= {1 \over 2} \int d\vec{k}_1 d\vec{k}_2 \pi_{k_1 k_2} \pi_{k_1 k_2} \cr
&&+{1 \over 8} \int d\vec{k}_1 d\vec{k}_2 \eta_{k_1 k_2} ({\psi^0}_{k_1}^{-1} + {\psi^0}_{k_2}^{-1})^2 \eta_{k_1 k_2}
\end{eqnarray}
from which one reads off the frequencies
\begin{equation}
\omega_{k_1 k_2}={1 \over 2}{\psi^0}_{k_1}^{-1}+{1 \over 2}{\psi^0}_{k_2}^{-1}=\sqrt{{\vec{k}_1}^2}+\sqrt{{\vec{k}_2}^2}.
\end{equation}
To summarize, the quadratic Hamiltonian and momentum can be written in use of bi-local fields as
\begin{eqnarray}
H^{(2)}&=&\int d\vec{x} d\vec{y} \Psi^\dagger (\vec{x},\vec{y})\Bigl( \sqrt{-\bigtriangledown_x^2}+\sqrt{-\bigtriangledown_y^2} \Bigr) \Psi(\vec{x},\vec{y}), \cr
P^{(2)}&=&\int d\vec{x} d\vec{y} \Psi^\dagger (\vec{x},\vec{y})(\bigtriangledown_x+\bigtriangledown_y) \Psi(\vec{x},\vec{y}).
\end{eqnarray}
In the light-cone quantization, we have the quadratic Hamiltonian
\begin{eqnarray}
&&P^{-(2)}=H^{(2)}+P^{(2)} \cr
&&=\int dx_1^- dx_2^- d\vec{x}_1 d\vec{x}_2 \Psi^\dagger \Bigl(-{\bigtriangledown_1^2 \over 2p_1^+}-{\bigtriangledown_2^2 \over 2p_2^+}\Bigr) \Psi.
\end{eqnarray}
Here $\Psi(x^+;x_1^-,x_2^-;\vec{x}_1,\vec{x}_2)$ is a bi-local field where $1,2$ refer to the two space points.

\section{Conformal transformations of the collective fields}

Our goal is to demonstrate that the collective field contains all the necessary information and is in a one-to-one map with the physical fields of the higher-spin theory in AdS$_4$. For this comparison to be done it is advantageous to work in the light-cone gauge, where the physical degrees of freedom of a gauge theory are most transparent \cite{Metsaev:1999ui, Metsaev:2000yu}. Our strategy is to compare directly the action of the conformal group of the $d=3$ field theory with that of the Anti de Sitter higher spin field. This comparison is similar to the study in D-brane case and ${\cal N}=4$ Super Yang-Mills theory performed in \cite{Jevicki:1998ub}. In this direct comparison we will see that as expected we have very different set of space-time variables and a different realization of SO(2,3). The number of canonical variables however will be shown to be identical and one can search for a (canonical) transformation to establish a one-to-one relation between the two representations.

One can work out the conformal transformations in light-cone notation ($x^+=t$) for any dimension $d$. As for the linear momenta, we have
\begin{eqnarray}
P^-&=&H=\int d\vec{x} \Bigl( -{1 \over 2}(\partial_i \phi)^2 \Bigr), \cr
P^+&=&\int d\vec{x} \Bigl( \pi^2 \Bigr), \cr
P^i&=&\int d\vec{x} \Bigl( \pi \partial_i \phi \Bigr),
\end{eqnarray}
where $\pi=\partial^+ \phi$ is the conjugate momentum and $i$ is the transverse index (for the specific case when $d=3$, the index $i$ runs over a single value). Similarly, for Lorentz transformations, the conserved charges are
\begin{eqnarray}
M^{+-}&=&t H-\int d\vec{x} \Bigl( x^- \pi^2 \Bigr), \cr
M^{+i}&=&\int d\vec{x} \Bigl( t \pi \partial_i \phi-x^i \pi^2 \Bigr), \cr
M^{-i}&=&\int d\vec{x} \Bigl( x^- \pi \partial_i \phi-x^i~{\cal H} \Bigr), \cr
M^{ij}&=&\int d\vec{x} \Bigl( x^i \pi \partial_j \phi-x^j \pi \partial_i \phi \Bigr).
\end{eqnarray}
The Dilatation operator takes the form
\begin{equation}
D=t H+\int d\vec{x} \Bigl( \pi(d_\phi+x^i \partial_i)\phi+x^- \pi^2 \Bigr),
\end{equation}
where $d_\phi={d-2 \over 2}$ is the scaling dimension of the $\phi$ field. The special conformal generators are
\begin{eqnarray}
K^-&=&\int d\vec{x} \Bigl( x^- ~{\cal D}-{1 \over 2}(2t x^- +x^j x_j){\cal H}-{1 \over 2}d_\phi \phi^2 \Bigr), \cr
K^+&=&t D-\int d\vec{x} \Bigl( {1 \over 2}(2t x^- + x^j x_j) \pi^2 \Bigr), \cr
K^i&=&\int d\vec{x} \Bigl( x^i ~{\cal D}-{1 \over 2}(2t x^- +x^j x_j)\pi \partial_i \phi \Bigr),
\end{eqnarray}
where ${\cal D}$ and ${\cal H}$ are the densities of these two operators.

The dynamical variables in the light-cone formulation are $(x^-,x^i)$. The momentum conjugate to $x^-$ is $p^+$. In the massless case, the energy can be expressed as
\begin{equation}
p^-=-{p^i p^i \over 2p^+}.
\end{equation}
To define the mode expansion, we perform a Fourier transform of the fields $\phi(x^-,x^i)$ and $\pi(x^-,x^i)$ along the $x^-$ direction. The creation and annihilation operators are defined in terms of
\begin{eqnarray}
&&\phi(x^-,x^i)=\int_0^\infty {dp^+ \over \sqrt{2\pi}}{1 \over \sqrt{2p^+}} \cr
&&\Bigl( a(p^+,x^i) e^{i p^+ x^-}+a^\dagger (p^+,x^i) e^{-i p^+ x^-} \Bigr), \\
&&\pi(x^-,x^i)=-i \int_0^\infty {dp^+ \over \sqrt{2\pi}} \sqrt{p^+ \over 2} \cr
&&\Bigl( a(p^+,x^i) e^{i p^+ x^-}-a^\dagger (p^+,x^i) e^{-i p^+ x^-} \Bigr).
\end{eqnarray}
The actions of linear momenta now take the form
\begin{eqnarray}
P^-: \quad \delta a(p^+,x^i)&=&{\partial_i^2 \over 2p^+} a(p^+,x^i), \cr
P^+: \quad \delta a(p^+,x^i)&=&p^+ a(p^+,x^i), \cr
P^i: \quad \delta a(p^+,x^i)&=&i\partial_i a(p^+,x^i).
\end{eqnarray}
For the Lorentz generators, one has
\begin{eqnarray}
M^{+-}: \quad \delta a(p^+,x^i)&=&\Bigl( t{\partial_i^2 \over 2p^+}-i\sqrt{p^+}{\partial \over \partial p^+}\sqrt{p^+} \Bigr) \cr
& & a(p^+,x^i), \cr
M^{+i}: \quad \delta a(p^+,x^i)&=&\Bigl( i t \partial_i -x^i p^+ \Bigr)a(p^+,x^i), \cr
M^{-i}: \quad \delta a(p^+,x^i)&=&\Bigl( -\partial_i {\partial \over \partial p^+}-{\partial_j x^i \partial_j \over 2p^+} \Bigr) a(p^+,x^i), \cr
M^{ij}: \quad \delta a(p^+,x^i)&=&\Bigl( i x^i \partial_j-i x^j \partial_i \Bigr) a(p^+,x^i).
\end{eqnarray}
and the Dilatation operator
\begin{eqnarray}
D: \quad \delta a(p^+,x^i)&=&\Bigl( t{\partial_i^2 \over 2p^+}+i\Bigl[d_\phi+x^i \partial_i \cr
&+&\sqrt{p^+}{\partial \over \partial p^+}\sqrt{p^+}\Bigr]\Bigr) a(p^+,x^i).
\end{eqnarray}
Finally, for the special conformal generators
\begin{eqnarray}
K^-: \quad \delta a(p^+,x^i)&=&\Bigl\{ -{\partial_j x^i x^i \partial_j \over 4p^+}-\sqrt{p^+}{\partial \over \partial p^+}{\partial \over \partial p^+}\sqrt{p^+} \cr
&&- x^i \partial_i {\partial \over \partial p^+}-d_\phi {1 \over \sqrt{p^+}} {\partial \over \partial p^+} \sqrt{p^+} \Bigr\} \cr
&&a(p^+,x^i), \cr
K^+: \quad \delta a(p^+,x^i)&=&\Bigl\{ t^2 {\partial_i^2 \over 2p^+}+it(d_\phi+x^i \partial_i)-{1 \over 2} x^i x^i p^+ \Bigr\} \cr
&&a(p^+,x^i), \cr
K^i: \quad \delta a(p^+,x^i)&=&\Bigl\{ t {\partial_j x^i \partial_j \over 2p^+}+t \partial_i {\partial \over \partial p^+}-{i \over 2}x^j x^j \partial_i \cr
&&+i x^i \Bigl[d_\phi+x^j\partial_j+\sqrt{p^+}{\partial \over \partial p^+}\sqrt{p^+}\Bigr] \Bigr\} \cr
&&a(p^+,x^i).
\end{eqnarray}

We next deduce the transformation for the collective fields. In creation-annihilation form $A(x_1^-,x_2^-,\vec{x}_1,\vec{x}_2)=a(x_1^-,\vec{x}_1)a(x_2^-,\vec{x}_2)$, we have $\delta A(1,2)=\delta a(1) a(2)+a(1) \delta a(2)$ and any conformal generator
\begin{eqnarray}
G&=&\int dx_1^- dx_2^- d\vec{x}_1 d\vec{x}_2 A^\dagger \hat{g} A \cr
&=&\int dx_1^- dx_2^- d\vec{x}_1 d\vec{x}_2 A^\dagger (\hat{g}_1+\hat{g}_2) A.
\label{little1}
\end{eqnarray}
Denoting the conjugate momenta as $(p_1^+,p_2^+,p_1^i,p_2^i)$, we can write down the following generators
\begin{eqnarray}
\hat{p}^-&=&p_1^-+p_2^-=-\Bigl({p_1^i p_1^i \over 2p_1^+}+{p_2^i p_2^i \over 2p_2^+}\Bigr), \label{cft1} \\
\hat{p}^+&=&p_1^++p_2^+, \label{cft2} \\
\hat{p}^i&=&p_1^i+p_2^i, \label{cft3} \\
\hat{m}^{+-}&=&t\hat{p}^--x_1^- p_1^+-x_2^- p_2^+, \label{cft4} \\
\hat{m}^{+i}&=&t\hat{p}^i-x_1^i p_1^+ -x_2^i p_2^+, \label{cft5} \\
\hat{m}^{-i}&=&x_1^- p_1^i+x_2^- p^i_2+x^i_1{p^j_1 p_1^j \over 2p_1^+}+x^i_2{p^j_2 p_2^j \over 2p_2^+}, \label{cft6} \\
\hat{d}&=&t\hat{p}^-+x_1^- p_1^+ + x_2^- p_2^+ \cr
&&+x^i_1 p_1^i +x^i_2 p_2^i+ 2d_\phi, \label{cft7} \\
\hat{k}^-&=&x^i_1 x_1^i{p^j_1 p_1^j \over 4p_1^+}+x^i_2 x_2^i{p^j_2 p_2^j \over 4p_2^+} \cr
&&+x_1^-(x_1^- p_1^++x^i_1 p_1^i +d_\phi) \cr
&&+x_2^-(x_2^- p_2^++x^i_2 p_2^i+d_\phi), \label{cft8} \\
\hat{k}^+&=&t^2 \hat{p}^-+t(x^i_1 p_1^i+x^i_2 p_2^i+2d_\phi) \cr
&&-{1 \over 2}x^i_1 x_1^i p_1^+ -{1 \over 2}x^i_2 x_2^i p_2^+, \label{cft9} \\
\hat{k}^i&=&-t \Bigl(x_1^i{p^j_1 p_1^j \over 2p_1^+}+x^i_2{p^j_2 p_2^j \over 2p_2^+}+x_1^- p^i_1+x_2^- p^i_2\Bigr) \cr
&&-{1 \over 2}x^j_1 x_1^j p_1^i-{1 \over 2} x^j_2 x_2^j p_2^i \cr
&&+x_1^i(x_1^- p_1^++x_1^j p_1^j+d_\phi) \cr
&&+x_2^i(x_2^- p_2^++x_2^j p_2^j+d_\phi). \label{cft10}
\end{eqnarray}

\section{Mapping to AdS$_4$}

The correspondence introduced in \cite{Klebanov:2002ja} is specific for $CFT_3 \leftrightarrow AdS_4$. We will from now on consider the case of $d=3$ for the vector model. In the light-cone notation, there is only one transverse dimension $x^i=x$ and $x^\mu=(x^+,x^-,x)$.

The AdS$_4$ spacetime coordinates in the light-cone notation ($x^+=t$) are denoted with the Poincar\'e metric
\begin{equation}
ds^2={2 dt dx^-+dx^2+dz^2 \over z^2}.
\end{equation}
The lowercase transverse index $i=1$ denotes $x$ only, while the uppercase transverse index $I=(1,2)$ denotes $(x,z)$. In AdS$_4$ higher-spin theory, the generators were worked out by Metsaev in \cite{Metsaev:1999ui} which we now summarize. 

\subsection{Conformal generators from higher-spin theory}

The four-dimensional case has the unique property that, after fixing light-cone gauge \cite{Bengtsson:1986kh}, the only physical states are the $\pm s$ helicity states \cite{Bengtsson:1983pd}. Let us now explain how to fix the light-cone gauge. Starting from the covariant notation
\begin{equation}
\vert \Phi \rangle=\sum_{s=1}^\infty \Phi^{\mu_1...\mu_s}a^\dagger_{\mu_1}...a^\dagger_{\mu_s} \vert 0 \rangle.
\end{equation}
where $\mu=(0,1,z,3)$ in the case of AdS$_4$, one fixes the light-cone gauge in two steps. First, we drop the oscillators $a^\pm=a^0 \pm a^3$ and keep only the transverse oscillators $a^I,a^{\dagger J}$ including the $z$ component. The oscillators satisfy the commutators
\begin{equation}
[a^I,a^{\dagger J}]=\delta^{IJ}, \quad [a^I,a^J]=[a^{\dagger I},a^{\dagger J}]=0.
\end{equation}
The spin matrix of the Lorentz algebra now takes the form
\begin{equation}
M^{IJ}=a^{\dagger I} a^J-a^{\dagger J} a^I.
\label{spinmatrix}
\end{equation}
The next step is to impose a further constraint
\begin{equation}
T \vert \Phi \rangle=0, \qquad T=a^I a^I
\end{equation}
so that only two components will survive. With the complex oscillators
\begin{eqnarray}
\alpha={1 \over \sqrt{2}}(a_1+i a_2), &\qquad& \alpha^\dagger={1 \over \sqrt{2}}(a_1^\dagger+i a_2^\dagger), \\
\bar{\alpha}={1 \over \sqrt{2}}(a_1-i a_2), &\qquad& {\bar{\alpha}}^\dagger={1 \over \sqrt{2}}(a_1^\dagger-i a_2^\dagger),
\end{eqnarray}
we find the simple expansion for $\vert \Phi \rangle$
\begin{equation}
\vert \Phi \rangle=\sum_{\lambda=1}^\infty \Bigl( \Phi_{(\lambda)} (\bar{\alpha}^\dagger)^\lambda+\bar{\Phi}_{(\lambda)}(\alpha^\dagger)^\lambda \Bigr) \vert 0 \rangle.
\end{equation}
This expansion obviously satisfies the constraint
\begin{equation}
T \vert \Phi \rangle=0, \qquad T=\bar{\alpha} \alpha.
\end{equation}
The spin matrix
\begin{equation}
M=\alpha^\dagger \bar{\alpha}-{\bar{\alpha}}^\dagger \alpha
\end{equation}
also reduces to (\ref{spinmatrix}).

In four dimensions, the only non-vanishing spin matrix is $M^{xz}$. One can represent $\alpha=e^{i\theta}$, $\bar{\alpha}=e^{-i\theta}$. In a coherent basis, the operator $M^{xz}$ becomes ${\partial \over \partial \theta}$. Then we have $\Phi(x^\mu,z,\theta)$ or in light-cone notation $\Phi(x^+,x^-,x,z;\theta)$. The generators can be written as
\begin{equation}
G=\int dx^- dx dz d\theta ~ \bar{\Phi} \hat{g} \Phi.
\label{little2}
\end{equation}
Denoting the conjugate momenta as $(p^+,p^x,p^z,p^\theta)$, one has \cite{Metsaev:1999ui}
\begin{eqnarray}
\hat{p}^-&=&-{p^x p^x+p^z p^z \over 2p^+}, \label{ads1} \\
\hat{p}^+&=&p^+, \label{ads2} \\
\hat{p}^x&=&p^x, \label{ads3} \\
\hat{m}^{+-}&=&t\hat{p}^--x^- p^+, \label{ads4} \\
\hat{m}^{+x}&=&t p^x-x p^+, \label{ads5} \\
\hat{m}^{-x}&=&x^- p^x-x \hat{p}^-+{p^\theta p^z \over p^+}, \label{ads6} \\
\hat{d}&=&t\hat{p}^-+x^- p^++x p^x+z p^z+d_a, \label{ads7} \\
\hat{k}^-&=&-{1 \over 2}(x^2+z^2)\hat{p}^-+x^-(x^-p^++x p^x+z p^z+d_a) \cr
&&+{1 \over p^+}\bigl((x p^z-z p^x)p^\theta+(p^\theta)^2\bigr), \label{ads8} \\
\hat{k}^+&=&t^2\hat{p}^-+t(x p^x+z p^z+d_a)-{1 \over 2}(x^2+z^2)p^+, \label{ads9} \\
\hat{k}^x&=&t(x \hat{p}^--x^- p^x-{p^\theta p^z \over p^+})+{1 \over 2}(x^2-z^2) p^x \cr
&&+x(x^- p^++z p^z+d_a)+z p^\theta, \label{ads10}
\end{eqnarray}
where the scaling dimension $d_a=1$ in the case of AdS$_4$.

\subsection{The map: canonical transformation}

We will now show how the two pictures are related by a canonical transformation. At this point, we will give the classical transformation (it can be specified in its full quantum version also). So in what follows we do not compare terms with $d_\phi$ which will receive quantum corrections (due to ordering).

By relating (\ref{cft2}-\ref{cft5}) to (\ref{ads2}-\ref{ads5}), one can easily solve for
\begin{eqnarray}
x^-&=&{x_1^- p_1^++x_2^- p_2^+ \over p_1^++p_2^+}, \\
p^+&=&p_1^+ + p_2^+, \label{momentum1} \\
x&=&{x_1 p_1^++x_2 p_2^+ \over p_1^++p_2^+}, \\
p^x&=&p_1+p_2. \label{momentum2}
\end{eqnarray}
From (\ref{ads1},\ref{ads6},\ref{ads7},\ref{ads9}), we get
\begin{eqnarray}
z^2&=&{(x_1-x_2)^2 p_1^+ p_2^+ \over (p_1^++p_2^+)^2}, \label{eqn1} \\
p^z p^z&=&{(p_1 p_2^+-p_2 p_1^+)^2 \over p_1^+ p_2^+}, \label{eqn2} \\
z p^z&=&{(x_1-x_2)(p_1 p_2^+-p_2 p_1^+) \over (p_1^++p_2^+)}, \label{eqn3} \\
p^\theta p^z&=&(x_1^--x_2^-)(p_1 p_2^+-p_2 p_1^+) \cr
&&+(x_1-x_2)\Bigl({p_2^+ (p_1)^2 \over 2p_1^+}-{p_1^+(p_2)^2 \over 2p_2^+}\Bigr). \label{eqn4}
\end{eqnarray}
The solution to (\ref{eqn1}-\ref{eqn4}) can be written as
\begin{eqnarray}
z&=&{(x_1-x_2)\sqrt{p_1^+ p_2^+} \over p_1^+ + p_2^+}, \\
p^z&=&\sqrt{p_2^+ \over p_1^+} p_1-\sqrt{p_1^+ \over p_2^+} p_2, \label{momentum3} \\
p^\theta&=&\sqrt{p_1^+ p_2^+}(x_1^--x_2^-) \cr
&&+{x_1-x_2 \over 2}\Bigl(\sqrt{p_2^+ \over p_1^+} p_1+\sqrt{p_1^+ \over p_2^+} p_2\Bigr).
\end{eqnarray}
A nontrivial check of the consistency is given by comparing (\ref{ads8},\ref{ads10}) with (\ref{cft8},\ref{cft10}).

We now turn to the construction of $\theta$. The condition that $\theta$ Poisson commutes with $p^x$ implies $\theta$ is a function of $x_1-x_2$ and the condition that $\theta$ Poisson commutes with $p^+$ implies that $\theta$ is a function of $x_1^- -x_2^-$. Requiring that $\theta$ Poisson commutes with $x^-$, $x$, $z$ and $p^z$ as well as $\theta$ and $p^\theta$ Poisson commute to give 1 we obtain
\begin{equation}
\theta = 2\arctan\sqrt{p_2^+\over p_1^+}.
\label{momentum4}
\end{equation}

An important consistency check on the correctness of the map that we have constructed is that all the Poission brackets of the derived variables (like $z$ and $p^z$ etc.) take the canonical form with distinct canonical sets commuting with each other. One can confirm the Poisson brackets
\begin{equation}
\{x^-,p^+\}=\{x,p^x\}=\{z,p^z\}=1
\end{equation}
and others vanish. 

Finally, as a consequence of the above map it follows that the wave equation in the collective picture has a map \cite{deMelloKoch:1996mj} to the wave equation of higher-spin gravity in four-dimensional AdS background. This follows from the little generators (\ref{cft1}) and (\ref{ads1}) coinciding after the canonical transformation. The canonical transformation can be understood as a point transformation in the momentum space (if we interpret $\theta$ as momentum (\ref{momentum4}), the other momenta are given by (\ref{momentum1},\ref{momentum2},\ref{momentum3})). Consequently, the transformation between the higher-spin field and bi-local field is simple in momentum space 
\begin{eqnarray}
&&\Phi(x^-,x,z,\theta)=\int dp^+ dp^x dp^z e^{i (x^- p^++x p^x+z p^z)} \cr
&&\int dp_1^+ dp_2^+ dp_1 dp_2 \delta(p_1^+ + p_2^+ - p^+)\delta(p_1+p_2-p^x) \cr
&&\delta\Bigl(p_1 \sqrt{p_2^+ / p_1^+} -p_2 \sqrt{p_1^+ / p_2^+}-p^z\Bigr) \cr
&&\delta\bigl(2\arctan\sqrt{p_2^+ / p_1^+}-\theta\bigr) \tilde{\Psi}(p_1^+,p_2^+,p_1,p_2)
\end{eqnarray}
where $\tilde{\Psi}(p_1^+,p_2^+,p_1,p_2)$ is the Fourier transform of the bi-local field $\Psi(x_1^-,x_2^-,x_1,x_2)$.

\section{Conclusion / Origin of the Extra Dimension}

The main contribution of this paper is an explicit one-to-one map between the collective field (in the case of the O(N) vector model) and the field of higher-spin gravity in 4D AdS space-time. This map is defined by the canonical transformation which establishes the relationship between the coordinates of the bi-local collective field and the coordinates of the AdS$_4$ space-time plus spin variables. The map is one to one, in particular the most telling formula is the one for the extra radial coordinate of AdS space-time
$$
z={(x_1-x_2)\sqrt{p_1^+ p_2^+} \over p_1^+ + p_2^+}.
$$

Here we have an explicit expression, in terms of the collective coordinates contained in the bi-local field. The physical picture for this extra dimension is much like the (collective) coordinates of solitons, which are contained in the field itself but are nontrivial to exhibit. Their origin is again through a canonical map from the existing field degrees of freedom. Naturally, if the boundary conditions are too restrictive then these degrees will be absent. In more recent phenomenological studies of scattering processes in QCD, a dipole picture \cite{Brower:2006ea} was used which can have a relation to the construction presented. It is interesting to confront this collective mechanism for the emerging dimension with other viewpoints such as holographic \cite{Akhmedov:1998vf}, Feynman diagrams \cite{Gopakumar:2003ns} and stochastic quantization \cite{Jevicki:1993rr}.

Returning to future issues we have the following. The collective field theory gives a bulk Hamiltonian representation for the higher-spin gravity. It specifies an infinite set of bulk interacting vertices, which can be explicitly evaluated. These can be compared with the higher spin approaches, in particular Vasiliev's and we expect to find agreement. This comparison is presently being performed. It is also interesting to consider various canonical gauge fixings of Vasiliev's theory.

\begin{acknowledgments}

AJ would like to thank J. Avan, S. Das, T. Yoneya and C. I. Tan for discussions and interest in this work. He is also grateful to Prof. T. Takayanagi for hospitality at the IPMU, Kashiwa, Tokyo where part of this work was done. KJ would like to thank X. Yin for interesting discussions on this subject, M. A. Vasiliev for clarifying one of his papers and also I. Messamah for the discussion on gravity. The work of AJ and KJ is supported by the Department of Energy under contract DE-FG-02-91ER40688. The work of KJ is also supported by the Galkin fellowship at Brown University. RdMK is supported by the South African Research Chairs Initiative of the Department of Science and Technology and National Research Foundation.

\end{acknowledgments}

\end{document}